\documentclass[a4paper]{revtex4}
\usepackage{graphicx}
\usepackage{fancyhdr}
\usepackage{amsmath,amssymb}
\usepackage{color}

\begin{document}

\title{Higgs Couplings and their Implications for New Physics Scales}


\author{M. Muhlleitner}
\affiliation{Institute for Theoretical Physics, Karlsruhe Institute of
Technology, Wolfgang-Gaede-Str.1, 76131 Karlsruhe, Germany}

\begin{abstract}
In view of the absence of any direct sign of New Physics (NP) at the LHC,
the precise investigation of the Higgs properties becomes more and
more important in our quest for physics beyond the Standard
Model (SM). Coupling measurements play here an important role and not only
complement the reach of the LHC but, depending on the physics
scenarios, also allow for tests of NP scales beyond the ones
accessible at present colliders. In this context, various
representative scenarios beyond the SM will be reviewed. 
\end{abstract}

\maketitle

\thispagestyle{fancy}


%%%%%%%%%%%%%%%%%%%%%%%%%%%%%%%%%%
\section{Introduction}
With the discovery of the Higgs boson by the LHC experiments ATLAS
\cite{Aad:2012tfa} and CMS \cite{Chatrchyan:2012ufa} in 2012, a change
of paradigm has taken place. The Higgs boson is not target of experimental
research any more but now serves as tool in our search for NP and
hence in the understanding of nature 
\cite{Kramer:2015pea,Gouzevitch:2014pua}. While the observed Higgs
particle is in good 
agreement with SM expectations the experimental
uncertainties are still large enough to allow for interpretation in a
variety of NP models beyond the SM (BSM). Higgs couplings will play a
crucial role here \cite{Englert:2014uua,Muhlleitner:2014eaa}. As shown
in Table~\ref{tab:coupprec}, the precision on the Higgs couplings will
increase from at present several tens of percent to about 10\% at the 
high-luminosity LHC (HL-LHC) and to about 1\% at  future $e^+ e^-$ linear
colliders (LC), see \cite{Englert:2014uua} and references therein. 
%------------------------------
\begin{table}[b!]
\begin{center}
\begin{tabular}{l||c|c||c|c||c}
\hline
{\small coupl.}	    &  {\small LHC}  & {\small HL-LHC}	       & {\small LC}
& {\small HL-LC} &  {\small comb.}\\
\hline\hline
$hWW$		    &  0.09 & 0.08	       & 0.011	   & 0.006 & 0.005	    \\
$hZZ$		    &  0.11 & 0.08	       & 0.008	   & 0.005 & 0.004	    \\
$htt$		    &  0.15 & 0.12	       & 0.040	   & 0.017 & 0.015	    \\
$hbb$		    &  0.20 & 0.16	       & 0.023	   & 0.012 & 0.011	    \\
$h\tau\tau$	    &  0.11 & 0.09	       & 0.033	   & 0.017 & 0.015	    \\
\hline
$h\gamma\gamma$     &  0.20 & 0.15	       & 0.083	   & 0.035 & 0.024	    \\
$hgg$		    &  0.30 & 0.08	       & 0.054	   & 0.028 & 0.024	    \\
\hline
$h_{\scriptsize \mathrm{invis}}$   &	---  & ---		& 0.008     & 0.004 & 0.004	     \\
\hline
\end{tabular}
\vspace*{0.2cm}
\caption{Expected accuracy at the 68\% C.L.\ on fundamental and derived
  Higgs couplings, with the deviations defined as $g=g_{\scriptsize
    \mbox{SM}} [1\pm\Delta]$ with respect to the SM at the LHC/HL-LHC
  (luminosities 300 and 3000 fb$^{-1}$), LC/HL-LC (energies
  250+500~GeV / 250+500~GeV+1~TeV and luminosities 250+500 fb$^{-1}$ /
  1150+1600+2500 fb$^{-1}$), and in combined analyses of HL-LHC and HL-LC.
  For invisible Higgs decays the upper limit on the underlying
  couplings is given. Taken from \cite{Englert:2014uua}.}
\label{tab:coupprec}
\end{center}
\vspace*{-0.6cm}
\end{table}
%------------------------------

The deviations in the Higgs couplings from the SM values can be due to
various NP effects: The Higgs particle can mix with other
scalars, it can be a composite particle or new particles can alter the
couplings through loop contributions. Depending on the strength and
the type of the coupling between the Higgs boson and NP, the limits
obtained from the Higgs measurements can be more stringent than those
derived from direct searches, electroweak (EW) precision measurements
or flavour physics. In this way precision Higgs physics can be
sensitive to NP showing up at scales much higher than the one given by
the vacuum expectation value (VEV) $v$ and open a unique window to BSM
sectors, that have not been strongly constrained yet by the present data. 

%%%%%%%%%%%%%%%%%%%%%%%%%%%%%%%%%%
\section{The Effective Lagrangian Approach}
In the analysis of NP effects, the effective Lagrangian approach
makes it possible to study a large class of BSM models in
terms of a well defined quantum field theory. It does not allow,
however, to investigate effects arising from light particles or Higgs
decays into new non-SM particles. For a complete picture of BSM
effects in Higgs physics, therefore the analysis has to be
complemented by studies within specific BSM models that capture such
features. Some representative examples shall be presented in the
following sections.  

The effective Lagrangian approach is based on the assumption of a few
basic principles, like {\it e.g.}~SM gauge symmetries. Deviations from
the SM are parametrised by higher-dimensional operators, that are
suppressed by the typical NP scale $\Lambda$. Assuming for simplicity
the Higgs boson to be CP-even and the 
conservation of baryon and lepton numbers, the leading BSM effects for the
Higgs boson being part of a weak doublet are parametrised by 53
dimension-6 operators
\cite{Burges:1983zg,Leung:1984ni,Buchmuller:1985jz,Grzadkowski:2010es}.
Based on operator expansions the deviations from the SM couplings are
then estimated to be of the order $g = g_{\text{SM}} [1+\Delta]$, with
$\Delta = {\cal O}(v^2/\Lambda^2)$ and the characteristic 
scale $\Lambda$ assumed to be much 
larger than the VEV, $\Lambda \gg v$. Note, however, that this
estimate cannot be applied in case the underlying model violates the
decoupling theorem. Assuming experimental accuracies of $\Delta
=0.2...0.01$, sensitivities to scales of order $\Lambda \sim 550$~GeV
up to 2.5~TeV can be achieved. The lower scale reach is complementary to
direct LHC searches. The larger bound, however, exceeds the
direct search range of the LHC in general. If NP is due to loop
effects, an additional loop suppression factor has to be taken into
account, leading to $\Delta = v^2/(16 \pi^2 \Lambda^2)$. The scales
that can be probed are then much lower, for $\Delta = 0.02$ we
get $\Lambda \approx 140$~GeV. 

\begin{figure}[t]
\centering
\includegraphics[width=90mm]{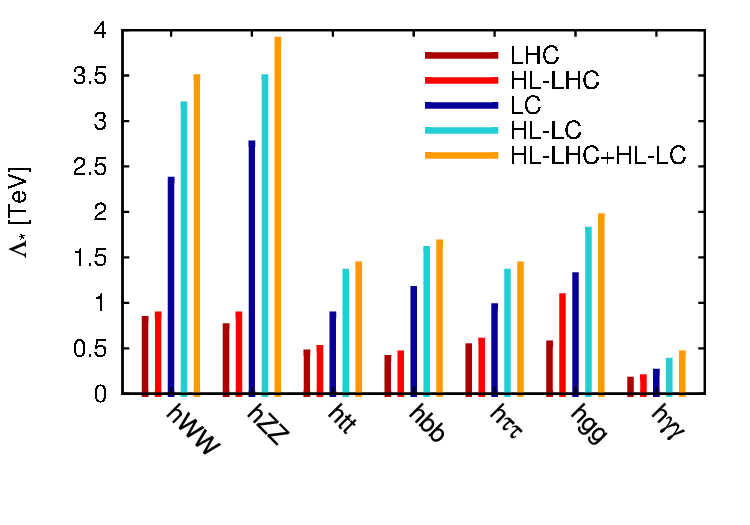}
\caption{Effective scales $\Lambda_\star$ to be probed at various
  collider options, based on the precisions given in
  Tab.~\ref{tab:coupprec}. For details, see \cite{Englert:2014uua}.} \label{fig:limits}
\end{figure}

In Fig.~\ref{fig:limits} we show the extracted limits on contributions
of the dimension-6 operators,
taking into account the precisions on the couplings given in
Table~\ref{tab:coupprec}. The limits have been derived with the
program {\tt sFitter} \cite{Lafaye:2009vr,Klute:2012pu,Plehn:2012iz}
after introducing the effective scales 
$\Lambda_\star$ that are obtained by factoring out from the operators
couplings and loop factors. Additionally in the loop-induced couplings
to the gluons and photons only the contributions from the contact
terms are included. The effects arising from loop terms are
disentangled already at the level of the input values $\Lambda$. 
The projected limits on $\Lambda_\star$ are summarized in
Table~\ref{tab:couplim}. As can be inferred from the table the
effective NP scales that can be probed in the Higgs sector range from
several hundred GeV to maximum values beyond a TeV. The bounds on new
particle masses $M$ exchanged in the Higgs vertex may, however, be reduced
significantly by small couplings, $M \sim \Lambda_\star
\sqrt{g^2/(16\pi^2)}$, where $g$ generically denotes the NP coupling.
%------------------------------
\begin{table}[h]
\begin{tabular}{|l||c|c||c|c||c|}
\hline
$\Lambda_\ast$ [TeV]     &  LHC  & HL-LHC & LC   & HL-LC & HL-LHC + HL-LC \\
\hline\hline
$hWW$                    &  0.82 & 0.87   & 2.35 & 3.18  & 3.48           \\
$hZZ$                    &  0.74 & 0.87   & 2.75 & 3.48  & 3.89           \\
\hline
$htt$                    &  0.45 & 0.50   & 0.87 & 1.34  & 1.42           \\
$hbb$                    &  0.39 & 0.44   & 1.15 & 1.59  & 1.66           \\
$h\tau\tau$              &  0.52 & 0.58   & 0.96 & 1.34  & 1.42           \\
\hline
$hgg$                    &  0.55 & 1.07   & 1.30 & 1.80  & 1.95           \\
$h\gamma\gamma$          &  0.15 & 0.18   & 0.24 & 0.36  & 0.44           \\
\hline
\end{tabular}
\caption{Effective NP scales $\Lambda_\ast$ extracted from the expected 
         accuracy on Higgs couplings at present and future colliders given in
         Table \ref{tab:coupprec}. Taken from \cite{Englert:2014uua}.} 
\label{tab:couplim}
\end{table}
%------------------------------

In case of non-linearly realized electroweak symmetry breaking (EWSB), the
most general effective Lagrangian at ${\cal O}(p^4)$ in a
derivative expansion, focusing on cubic terms with at least one
Higgs boson, assuming CP conservation and vector fields
coupling to conserved currents, is given by \cite{Contino:2010mh,Azatov:2012bz,Alonso:2012px,Buchalla:2013rka,Brivio:2013pma},
\begin{eqnarray} 
{\cal L} &=&
 \, \frac{1}{2} \partial_\mu h\ \partial^\mu h - \frac{1}{2} m_h^2 h^2 
- c_3 \, \frac{1}{6}\, \left(\frac{3m_h^2}{v}\right) h^3 - \sum_{\psi = u,d,l} m_{\psi^{(i)}} \, \bar\psi^{(i)}\psi^{(i)} \left( 1 + c_\psi \frac{h}{v} + \dots \right)
\nonumber \\[0.cm]
&& + m_W^2\,  W^+_\mu W^{-\, \mu} \left(1 + 2 c_W\, \frac{h}{v} +
  \dots \right) + \frac{1}{2} m_Z^2\,  Z_\mu Z^\mu \left(1 + 2 c_Z\, \frac{h}{v} + \dots \right) + \dots
\nonumber \\[0.3cm]
&& + \left(  \frac{\bar{c}_{WW} \,\alpha}{\pi} \,
  W_{\mu\nu}^+ W^{-\mu\nu}  + \frac{\bar{c}_{ZZ} \,\alpha}{2\pi}
  \, Z_{\mu\nu}Z^{\mu\nu} +  
\frac{\bar{c}_{Z\gamma} \, \alpha}{\pi}\, Z_{\mu\nu}
\gamma^{\mu\nu}   + \frac{\bar{c}_{\gamma\gamma} \,
\alpha}{2\pi}\, \gamma_{\mu\nu}\gamma^{\mu\nu} +
\frac{\bar{c}_{gg} \, \alpha_s}{12\pi}\, G_{\mu\nu}^aG^{a\mu\nu} \right)
\frac{h}{v} 
\nonumber \\[0.2cm]
&& +  \Big( \left(\bar{c}_{W\partial W}\, W^-_\nu D_\mu W^{+\mu\nu}+ h.c.\right)+\bar{c}_{Z\partial Z}\,  Z_\nu\partial_\mu Z^{\mu\nu}
+  \bar{c}_{Z\partial \gamma}\, Z_\nu\partial_\mu\gamma^{\mu\nu} \Big)\,
\frac{h}{v} + \dots \;,
\end{eqnarray}
where $\psi$ denotes the fermion fields, $\alpha$ the EW and
$\alpha_s$ the strong coupling. The EW and photon fields are described
by $Z,W$ and $\gamma$, the gluon fields by $G$, with $a$ being the
color index. The couplings $c_i,\bar{c}_i$ can take arbitrary values
and the Higgs boson $h$ need not be part of an electroweak
doublet. The couplings are truly independent of other parameters that
do not involve the Higgs boson. Applying the Lagrangian to the linear 
realization on the other hand, only 4 couplings between the Higgs
boson $h$ and the vector bosons $V$ are independent of the other EW
measurements \cite{Elias-Miro:2013mua,Pomarol:2013zra}. For a
discussion of the physics implications, {\it 
  cf.}~\cite{Contino:2013kra}. In the SM limit $c_i=1$ and
$\bar{c}_i = 0$. 

%%%%%%%%%%%%%%%%%%%%%%%%%%%%%%%%%%
\section{Composite Higgs Models}
In Composite Higgs Models a light Higgs boson arises as a pseudo
Nambu-Goldstone boson from a strongly-interacting sector
\cite{Dimopoulos:1981xc,Kaplan:1983fs,Banks:1984gj,Kaplan:1983sm,Georgi:1984ef,Georgi:1984af,Dugan:1984hq},
implying modified couplings compared to the SM. Such models are examples for
EWSB based on a strong dynamics. In~\cite{Giudice:2007fh} an effective
low-energy description of a Strongly Interacting Light Higgs Boson
(SILH) has been given, which can be viewed as first term of the
expansion in the compositeness parameter $\xi=\frac{v^2}{f^2}$, 
where $v\approx 246$~GeV is the VEV and $f$ the scale of the strong
dynamics. The SILH Lagrangian is applicable in the vicinity of the SM
limit, {\it i.e.}~$\xi \to 0$, but for larger values of $\xi$ a
resummation of the series in $\xi$ has to be performed. Explicit
models built in five-dimensional warped space provide such a
resummation: In the Minimal Composite Higgs Models (MCHM) discussed in
Refs.~\cite{Agashe:2004rs,Contino:2006qr} the global $SO(5)\times U(1)$
symmetry is broken down at 
the scale $f$ to $SO(4)\times U(1)$ on the infrared brane and to the
SM group $SU(2)_L \times U(1)_Y$ on the ultraviolet
brane. (For an MCHM implementing the antisymmetric
  representation ${\bf 10}$, see {\it e.g.}~\cite{Gillioz:2013pba}.) In these
models the Higgs coupling modifications can be described by one single
parameter, namely $\xi$. In the model, named MCHM4, of
\cite{Agashe:2004rs} the fermions are in the  spinorial representation
of $SO(5)$. Here all Higgs couplings are suppressed by the
same universal factor $(1-\xi)^{1/2}$. This case is covered by the
analysis of portal models. In MCHM5, the fermions are in the
fundamental representation of $SO(5)$, and the couplings to massive
gauge bosons $V=W,Z$ and the ones to fermions $f$ are modified by a
different coefficient with respect to the SM,
\begin{equation}
1 + \Delta_V = \sqrt{1-\xi} \approx 1-\frac{\xi}{2} \;, \quad 
1 + \Delta_f = \frac{1-2\xi}{\sqrt{1-\xi}} \approx 1 -\frac{3\xi}{2} \;,
\end{equation}
for $\xi \ll 1$. Table~\ref{tab:xilimits} shows the bounds on $\xi$,
respectively the scale $f$, derived by assuming the Higgs coupling
precisions given in Table~\ref{tab:coupprec}.
%------------------------------                                                
\begin{table}[t]
\begin{tabular}{|l||c|c||c|c||c|}
\hline 
$\xi$           & LHC    & HL-LHC & LC     & HL-LC & HL-LHC+HL-LC \\
\hline
universal       & 0.076  & 0.051  & 0.008  & 0.0052 & 0.0052      \\
non-universal   & 0.068  & 0.015  & 0.0023 & 0.0019 & 0.0019      \\    
\hline \hline
$f$ [TeV]       &        &        &        &        &             \\  
\hline
universal       & 0.89   & 1.09   & 2.82   & 3.41   & 3.41        \\
non-universal   & 0.94   & 1.98   & 5.13   & 5.65   & 5.65        \\\hline 
\end{tabular}
\caption{Derived bounds on the parameter $\xi = (v/f)^2$ and the Goldstone
             scale $f$ for various experimental set-ups. Taken from
             \cite{Englert:2014uua}.} 
\label{tab:xilimits}
\end{table}
%------------------------------ 

The computation of the Higgs boson decay widths
and branching ratios can be performed with the Fortran code {\tt
  eHDECAY}, \cite{Contino:2014aaa} which has implemented different
parametrisations 
of effective Lagrangians, the SILH approach, the non-linear
realization of EWSB and the composite Higgs models MCHM4 and
MCHM5. The program includes the most important
higher-order QCD effects and in case of the SILH and composite Higgs
parametrisation also the EW higher order corrections. The user
furthermore has the possibility to turn off these EW
corrections. 

%%%%%%%%%%%%%%%%%%%%%%%%%%%%%%%%%%
\section{The Two-Higgs-Doublet Model and the MSSM}
Two-Higgs-Doublet Models (2HDM) and the Minimal Supersymmetric
extension of the SM (MSSM) are examples of NP, where the Higgs
couplings are modified due to mixing effects. The 2HDM 
\cite{Lee:1973iz,Flores:1982pr,Gunion:1989we,Branco:2011iw,Gunion:2002zf}
belongs to the simplest extensions of 
the SM that allow to respect the experimentally measured $\rho$
parameter. The physical Higgs states are mixtures of the components of the
two doublets $\phi_1$ and $\phi_2$. The scalar potential reads
\begin{eqnarray}
V &=& m_{11} |\phi_1|^2 + m_{22}^2 |\phi_2|^2 - m_{12}^2(\phi_1^\dagger \phi_2 +
\text{h.c}) + \lambda_1 |\phi_1|^4 + \lambda_2 |\phi_2|^4 \nonumber \\
&&+ \lambda_3 |\phi_1|^2 |\phi_2|^2 + \lambda_4 |\phi_1^\dagger \phi_2|^2 
 + \frac{1}{2}\lambda_5 [(\phi_1^\dagger \phi_2)^2 + \text{h.c}]\,.
\label{eq:2hdmpot}
\end{eqnarray}
The neutral components of the Higgs doublets acquire VEVs $v_1$ and
$v_2$, with the ratio defined as $\tan\beta= v_2/v_1$. They add up
to $v^2=v_1^2 + v_2^2$. The 2HDM features five Higgs bosons after
EWSB, two neutral CP-even bosons $h^0$ and $H^0$, one CP-odd Higgs
$A^0$ and two charged states $H^\pm$. The Higgs couplings to the fermions are
different in specific realizations of the 2HDM. These arise by
demanding a natural suppression of flavour-changing neutral
currents, which is achieved by requiring that one type of fermions
couples only to one Higgs doublet. It can be assured by imposing a
global $\mathbb{Z}_2$ symmetry, under which $\phi_{1,2} \to \mp
\phi_{1,2}$. This symmetry has been assumed in the potential
Eq.~(\ref{eq:2hdmpot}), implying that all terms of the potential
include an even power of each of the Higgs fields. There are four
cases of possible couplings between the Higgs doublets and the
fermions, that depend on the $\mathbb{Z}_2$ charge assignment
\cite{Barger:1989fj}. 
\begin{figure}[b]
\centering
\includegraphics[width=65mm]{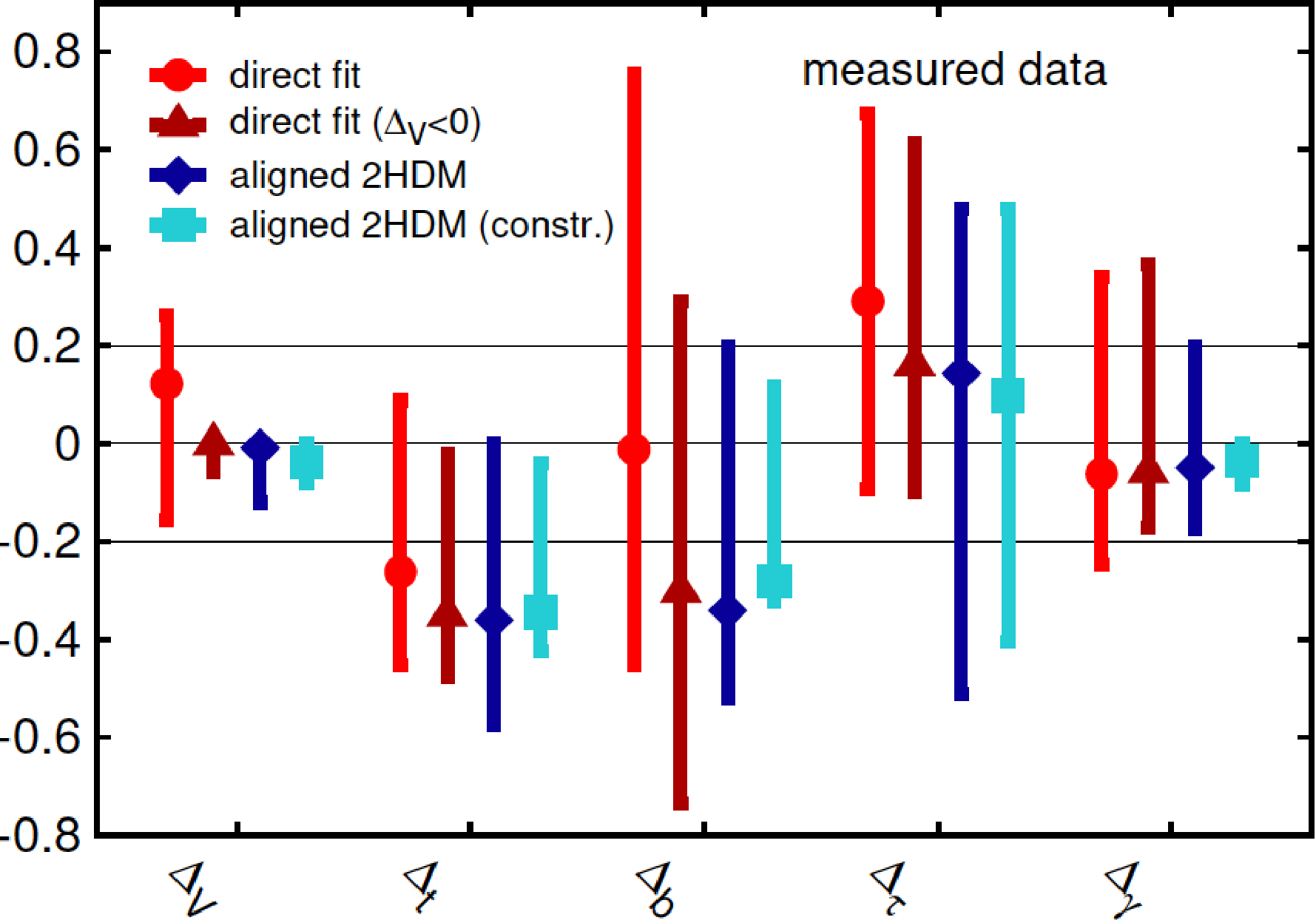}
\hspace*{1cm}
\includegraphics[width=72mm]{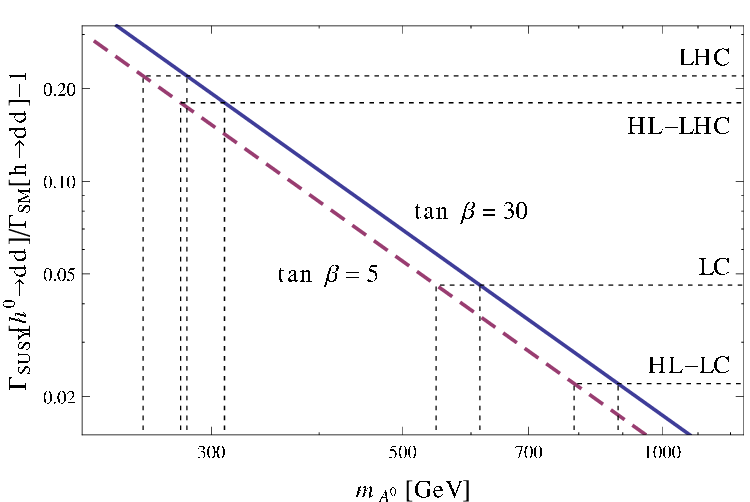}
\caption{Left: Fits to the weak scale couplings and to the aligned 2HDM
  in terms of the light Higgs couplings 
  based on the data from the 7 TeV run and an integrated luminosity of
  4.6-5.1~fb$^{-1}$ and the 8 TeV run with 12-21~fb$^{-1}$, at 68\% C.L.,
  ATLAS and CMS 
  combined \cite{Lopez-Val:2013yba}. Right: Deviation of the MSSM
  tree-level decay width into down-type 
  fermions from the SM value as function of $m_{A^0}$ for
  $\tan\beta=5$ and 30. For further details on $R_t$, see
  \cite{Englert:2014uua}. Shown are the expected
  limits at the LHC, HL-LHC, LC and HL-LC.} \label{fig:2hdm}
\end{figure}
%In type I all fermions couple only to $\phi_2$,
%in type II up- and down-type fermions couple to $\phi_2$ and $\phi_1$,
%respectively, in lepton-specific models the quarks couple to $\phi_2$
%and the leptons to $\phi_1$ and in flipped 2HDMs up-type quarks and
%leptons couple to $\phi_2$, whereas the down-type quarks couple to
%$\phi_1$. 
In the aligned 2HDM, the Yukawa couplings of the two Higgs
doublets are proportional to each other in flavour space. At tree
level the aligned 2HDM is determined by five free parameters,
including the mass of the charged Higgs boson, that 
contributes to the effective Higgs-photon coupling.
Figure~\ref{fig:2hdm} (left) compares the extracted free Higgs
couplings with the corresponding fit to the aligned 2HDM parameters,
translated into the SM  coupling deviations. For simplicity custodial
symmetry, {\it i.e.}~$\Delta_Z = \Delta_W \equiv \Delta V < 1$ has
been assumed. Note, that there are additional
constraints due to non-standard Higgs searches and EW precision
measurements as well as flavour constraints, in case the 2HDM is
realized. These are taken into account in the cyan bands, while they
have been ignored in the blue ones. The 2HDM has also been implemented
in the Fortran code {\tt HDECAY} 
\cite{Djouadi:1997yw,Djouadi:2006bz,Butterworth:2010ym} to
provide the Higgs decay widths and branching ratios including the
state-of-the-art higher order QCD corrections and the off-shell Higgs
decays \cite{Harlander:2013qxa}. 

The Higgs sector of the MSSM is a subgroup of the general 2HDM type-II
where the up- and down-type fermions couple to $\phi_2$ and $\phi_1$,
respectively. Furthermore, the quartic couplings are given in terms of
the $SU(2)_L$ and $U(1)_Y$ gauge couplings. In the decoupling limit, where
$H^0, A^0$ and $H^\pm$ are heavy and $h^0$ behaves SM-like, the partial
decay width of the latter into down-type fermions, normalized to the
SM, scales with the pseudoscalar and the $Z$ boson masses, $\beta$ and
the dominant supersymmetric radiative corrections $R_t$. 
%\begin{equation}
%\frac{\Gamma_{\text{SUSY}} (h^0 \to dd)}{\Gamma_{\text{SM}} (h\to dd)}
%\approx 1 - \frac{4 m_Z^2 \sin^2 \beta}{m_{A^0}^2} (\cos 2 \beta +
%R_t) \;.
%\end{equation}
%For the definition of $R_t$, see
%\cite{Englert:2014uua}. 
Figure~\ref{fig:2hdm} (right) shows the deviation of
this decay width in the MSSM from the SM as a function of $m_{A^0}$
for two different $\tan\beta$ values. It is proportional to the deviation in
the Higgs coupling squared, and depending on the coupling precision
achieved at the various colliders, limits on $m_{A^0}$ can be derived
for a fixed value of $\tan\beta$.
\begin{figure}[t]
\centering
\includegraphics[width=75mm]{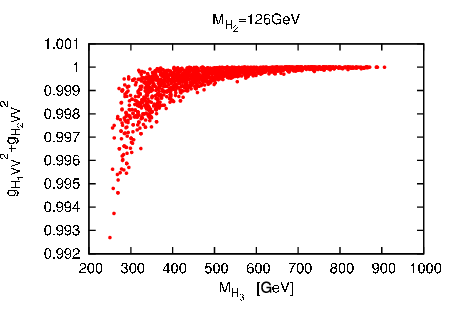}
\hspace*{1cm}
\includegraphics[width=75mm]{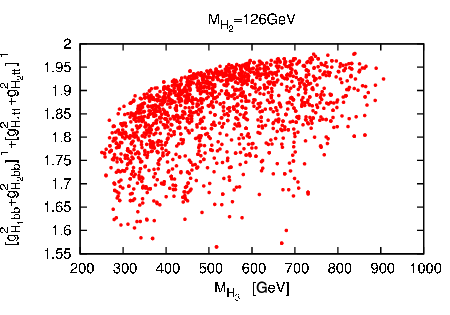}
\caption{Left: Sum of the $H_1$ and $H_2$ (SM-like) gauge
  couplings squared. Right: Sum of their inverse Yukawa couplings
  squared. From \cite{Englert:2014uua}.} \label{fig:nmssm}
\end{figure}

%%%%%%%%%%%%%%%%%%%%%%%%%%%%%%%%%%
\section{The NMSSM}
The NMSSM (for recent reviews, see
\cite{Maniatis:2009re,Ellwanger:2009dp}) includes an additional 
singlet superfield and features 7 Higgs bosons after EWSB,
three neutral CP-even ones $H_{1,2,3}$, two neutral CP-odd bosons
$A_{1,2}$ and two charged Higgs bosons $H^\pm$. Due to the large number of
parameters entering the tree-level Higgs sector, there are more
possibilities to achieve an NMSSM scenario compatible with the present
LHC data (see, {\it
  e.g.}~\cite{King:2012is,King:2012tr,King:2014xwa}). At the same time
it becomes more 
difficult to constrain a single parameter or a subset of the
parameters based on the coupling measurements alone. However, the
latter may allow to reveal if the possibly discovered new Higgs
particles belong to the MSSM or the NMSSM, in case only three have
been discovered and not all of them are CP-even. For the NMSSM this
would reveal itself in the violation of the coupling sum rules for the
scalar couplings to the gauge bosons,
\begin{equation}
\sum_{i=1}^3 g_{H_i VV}^2 = 1 \;,
\end{equation} 
and for the couplings to the top and bottom quarks,
\begin{equation}
\frac{1}{\sum_{i=1}^3 g_{H_itt}^2} + \frac{1}{\sum_{i=1}^3
  g_{H_ibb}^2} =1 \;.
\end{equation}
Figure~\ref{fig:nmssm} shows the scenarios with $H_2$ being SM-like
from a scan over the NMSSM parameter range, which are in accordance
with the LHC Higgs data. Assuming that only the two lightest CP-even
Higgs bosons have been discovered, the left plot shows as a function
of $M_{H_3}$ the violation of the vector coupling sum rule for $H_1$
and $H_2$ and the right plot the violation of the Yukawa coupling sum
rule. The sums can deviate by up to a factor of two in case of the
fermion couplings. While the precise determination of the Higgs
couplings will allow to distinguish the MSSM from the NMSSM it will be
difficult to deduce the mass of the unobserved third Higgs boson from
the pattern of the violation of the sum rules. The larger number of
parameters entering the Higgs sector does not allow to derive a unique
correlation between the coupling values and the scale of NP. In this
case a global scan has to be performed to pin down the underlying NP scale.

%%%%%%%%%%%%%%%%%%%%%%%%%%%%%%%%%%
\section{Conclusion}
The precise investigation of the Higgs properties opens a unique
window to NP scales beyond the direct reach of present colliders. It
has been shown for some archetypal BSM scenarios that Higgs precision
data can be sensitive to scales ranging from a few hundred GeV in
weakly coupled models up to multi-TeV scales for models based on
strong dynamics.

%%%%%%%%%%%%%%%%%%%%%%%%%%%%%%%%%%

\begin{acknowledgments}
I would like thank the organisers for the nice and fruitful workshop
and for the invitation to give the talk.
\end{acknowledgments}

\bigskip % extra skip inserted
%\begin{thebibliography}{99} % Use for 10-99 references
%\bibitem{DPF2009} http://www.dpf2009.wayne.edu/proceedings.php
%\bibitem{eConf} http://www.slac.stanford.edu/econf/
%\bibitem{templates-ref} http://www.slac.stanford.edu/econf/editors/eprint-template/instructions.html
%\bibitem{arXiv} http://arxiv.org/help
%
%\end{thebibliography}

%%%%%%%%%%%%%%%%%%%%%%%%%%%%%%%%%%%%%%%%%%%%%%%%%%%%%%%%%%
%\bibliographystyle{h-physrev}
%\bibliographystyle{plain}

%\bibliography{HPNP2015_muhlleitner}

\begin{thebibliography}{100}

\bibitem{Aad:2012tfa}
ATLAS Collaboration, G.~Aad {\em et~al.},
Phys.Lett. {\bf B716}, 1 (2012), 1207.7214.

\bibitem{Chatrchyan:2012ufa}
CMS Collaboration, S.~Chatrchyan {\em et~al.},
Phys.Lett. {\bf B716}, 30 (2012), 1207.7235.

\bibitem{Kramer:2015pea}
M.~Kramer and M.~Muhlleitner,
Nucl.Part.Phys.Proc. {\bf 261-262}, 246 (2015), 1501.06658.

\bibitem{Gouzevitch:2014pua}
M.~Gouzevitch, A.~Kaczmarska, M.~Muhlleitner, and K.~Turzynski,
PoS {\bf DIS2014}, 003 (2014).

\bibitem{Englert:2014uua}
C.~Englert {\em et~al.},
J.Phys. {\bf G41}, 113001 (2014), 1403.7191.

\bibitem{Muhlleitner:2014eaa}
M.~Muhlleitner,
(2014), 1410.5093.

\bibitem{Burges:1983zg}
C.~Burges and H.~J. Schnitzer,
Nucl.Phys. {\bf B228}, 464 (1983).

\bibitem{Leung:1984ni}
C.~N. Leung, S.~Love, and S.~Rao,
Z.Phys. {\bf C31}, 433 (1986).

\bibitem{Buchmuller:1985jz}
W.~Buchmuller and D.~Wyler,
Nucl.Phys. {\bf B268}, 621 (1986).

\bibitem{Grzadkowski:2010es}
B.~Grzadkowski, M.~Iskrzynski, M.~Misiak, and J.~Rosiek,
JHEP {\bf 1010}, 085 (2010), 1008.4884.

\bibitem{Lafaye:2009vr}
R.~Lafaye, T.~Plehn, M.~Rauch, D.~Zerwas, and M.~Duhrssen,
JHEP {\bf 0908}, 009 (2009), 0904.3866.

\bibitem{Klute:2012pu}
M.~Klute, R.~Lafaye, T.~Plehn, M.~Rauch, and D.~Zerwas,
Phys.Rev.Lett. {\bf 109}, 101801 (2012), 1205.2699.

\bibitem{Plehn:2012iz}
T.~Plehn and M.~Rauch,
Europhys.Lett. {\bf 100}, 11002 (2012), 1207.6108.

\bibitem{Contino:2010mh}
R.~Contino, C.~Grojean, M.~Moretti, F.~Piccinini, and R.~Rattazzi,
JHEP {\bf 1005}, 089 (2010), 1002.1011.

\bibitem{Azatov:2012bz}
A.~Azatov, R.~Contino, and J.~Galloway,
JHEP {\bf 1204}, 127 (2012), 1202.3415.

\bibitem{Alonso:2012px}
R.~Alonso, M.~Gavela, L.~Merlo, S.~Rigolin, and J.~Yepes,
Phys.Lett. {\bf B722}, 330 (2013), 1212.3305.

\bibitem{Buchalla:2013rka}
G.~Buchalla, O.~Catà, and C.~Krause,
Nucl.Phys. {\bf B880}, 552 (2014), 1307.5017.

\bibitem{Brivio:2013pma}
I.~Brivio {\em et~al.},
JHEP {\bf 1403}, 024 (2014), 1311.1823.

\bibitem{Elias-Miro:2013mua}
J.~Elias-Miro, J.~Espinosa, E.~Masso, and A.~Pomarol,
JHEP {\bf 1311}, 066 (2013), 1308.1879.

\bibitem{Pomarol:2013zra}
A.~Pomarol and F.~Riva,
JHEP {\bf 1401}, 151 (2014), 1308.2803.

\bibitem{Contino:2013kra}
R.~Contino, M.~Ghezzi, C.~Grojean, M.~Muhlleitner, and M.~Spira,
JHEP {\bf 1307}, 035 (2013), 1303.3876.

\bibitem{Dimopoulos:1981xc}
S.~Dimopoulos and J.~Preskill,
Nucl.Phys. {\bf B199}, 206 (1982).

\bibitem{Kaplan:1983fs}
D.~B. Kaplan and H.~Georgi,
Phys.Lett. {\bf B136}, 183 (1984).

\bibitem{Banks:1984gj}
T.~Banks,
Nucl.Phys. {\bf B243}, 125 (1984).

\bibitem{Kaplan:1983sm}
D.~B. Kaplan, H.~Georgi, and S.~Dimopoulos,
Phys.Lett. {\bf B136}, 187 (1984).

\bibitem{Georgi:1984ef}
H.~Georgi, D.~B. Kaplan, and P.~Galison,
Phys.Lett. {\bf B143}, 152 (1984).

\bibitem{Georgi:1984af}
H.~Georgi and D.~B. Kaplan,
Phys.Lett. {\bf B145}, 216 (1984).

\bibitem{Dugan:1984hq}
M.~J. Dugan, H.~Georgi, and D.~B. Kaplan,
Nucl.Phys. {\bf B254}, 299 (1985).

\bibitem{Giudice:2007fh}
G.~Giudice, C.~Grojean, A.~Pomarol, and R.~Rattazzi,
JHEP {\bf 0706}, 045 (2007), hep-ph/0703164.

\bibitem{Agashe:2004rs}
K.~Agashe, R.~Contino, and A.~Pomarol,
Nucl.Phys. {\bf B719}, 165 (2005), hep-ph/0412089.

\bibitem{Contino:2006qr}
R.~Contino, L.~Da~Rold, and A.~Pomarol,
Phys.Rev. {\bf D75}, 055014 (2007), hep-ph/0612048.

\bibitem{Gillioz:2013pba}
M.~Gillioz, R.~Grober, A.~Kapuvari, and M.~Muhlleitner,
JHEP {\bf 1403}, 037 (2014), 1311.4453.

\bibitem{Contino:2014aaa}
R.~Contino, M.~Ghezzi, C.~Grojean, M.~Muhlleitner, and M.~Spira,
Comput.Phys.Commun. {\bf 185}, 3412 (2014), 1403.3381.

\bibitem{Lee:1973iz}
T.~Lee,
Phys.Rev. {\bf D8}, 1226 (1973).

\bibitem{Flores:1982pr}
R.~A. Flores and M.~Sher,
Annals Phys. {\bf 148}, 95 (1983).

\bibitem{Gunion:1989we}
J.~F. Gunion, H.~E. Haber, G.~L. Kane, and S.~Dawson,
Front.Phys. {\bf 80}, 1 (2000).

\bibitem{Branco:2011iw}
G.~Branco {\em et~al.},
Phys.Rept. {\bf 516}, 1 (2012), 1106.0034.

\bibitem{Gunion:2002zf}
J.~F. Gunion and H.~E. Haber,
Phys.Rev. {\bf D67}, 075019 (2003), hep-ph/0207010.

\bibitem{Barger:1989fj}
V.~D. Barger, J.~Hewett, and R.~Phillips,
Phys.Rev. {\bf D41}, 3421 (1990).

\bibitem{Lopez-Val:2013yba}
D.~Lopez-Val, T.~Plehn, and M.~Rauch,
JHEP {\bf 1310}, 134 (2013), 1308.1979.

\bibitem{Djouadi:1997yw}
A.~Djouadi, J.~Kalinowski, and M.~Spira,
Comput.Phys.Commun. {\bf 108}, 56 (1998), hep-ph/9704448.

\bibitem{Djouadi:2006bz}
A.~Djouadi, M.~Muhlleitner, and M.~Spira,
Acta Phys.Polon. {\bf B38}, 635 (2007), hep-ph/0609292.

\bibitem{Butterworth:2010ym}
J.~Butterworth {\em et~al.},
(2010), 1003.1643.

\bibitem{Harlander:2013qxa}
R.~Harlander, M.~Muhlleitner, J.~Rathsman, M.~Spira, and O.~Stal,
(2013), 1312.5571.

\bibitem{Maniatis:2009re}
M.~Maniatis,
Int.J.Mod.Phys. {\bf A25}, 3505 (2010), 0906.0777.

\bibitem{Ellwanger:2009dp}
U.~Ellwanger, C.~Hugonie, and A.~M. Teixeira,
Phys.Rept. {\bf 496}, 1 (2010), 0910.1785.

\bibitem{King:2012is}
S.~King, M.~Muhlleitner, and R.~Nevzorov,
Nucl.Phys. {\bf B860}, 207 (2012), 1201.2671.

\bibitem{King:2012tr}
S.~King, M.~Muhlleitner, R.~Nevzorov, and K.~Walz,
Nucl.Phys. {\bf B870}, 323 (2013), 1211.5074.

\bibitem{King:2014xwa}
S.~King, M.~Muhlleitner, R.~Nevzorov, and K.~Walz,
Phys.Rev. {\bf D90}, 095014 (2014), 1408.1120.

\end{thebibliography}
%%%%%%%%%%%%%%%%%%%%%%%%%%%%%%%%%%%%%%%%%%%%%%%%%%%%%%%%%%

\end{document}